\documentclass[useAMS,usenatbib]{mn2e}
\usepackage{epsfig}
\newcommand{\beq}{\begin{equation}}
\newcommand{\eeq}{\end{equation}}

\newdimen\hssize
\newdimen\hdsize 
\title[Clustering of narrow-line AGN]
{The clustering of narrow-line AGN in the local Universe}
\author[Li, Kauffmann, Wang, White \& Heckman]
{Cheng Li$^{1,2,3}$\thanks{E-mail: leech@ustc.edu.cn}, 
 Guinevere Kauffmann$^1$, Lan Wang $^{4,1}$, Simon D. M. White$^1$,
\newauthor Timothy M. Heckman$^5$, Y. P. Jing$^3$ \\
$^1$ Max-Planck Institut f\"{u}r Astrophysik, 
     D-85748 Garching, Germany \\
$^2$ Center for Astrophysics, University of Science 
     and Technology of China, Hefei, Anhui 230026, China \\
$^3$ The Partner Group of MPI f\"ur Astrophysik, 
     Shanghai Astronomical Observatory,
     Nandan Road 80, Shanghai 200030, China \\
$^4$ Department of Astronomy, Peking University, 
     Beijing 100871, China \\
$^5$ Department of Physics and Astronomy, 
     Johns Hopkins University, Baltimore, MD 21218}

\begin{document}

\date{Accepted ........ Received ........; in original form ........}

\pagerange{\pageref{firstpage}--\pageref{lastpage}} \pubyear{2006}

\maketitle

\label{firstpage}

\begin {abstract}
We have analyzed the clustering of $\sim 90,000$ narrow-line AGN drawn
from the  Data Release 4 (DR4)  of the Sloan Digital  Sky Survey.  Our
analysis addresses the following questions: a) How do the locations of
galaxies within the large-scale  distribution of dark matter influence
ongoing accretion onto their central  black holes?  b) Is AGN activity
triggered by interactions or mergers between galaxies?  We compute the
cross-correlation between AGN and a reference sample of galaxies drawn
from  the DR4.   We compare  this to  results for  control  samples of
inactive  galaxies matched simultaneously  in redshift,  stellar mass,
concentration, velocity  dispersion and mean stellar  age, as measured
by  the 4000  \AA\  break strength.   We  also compare  near-neighbour
counts around AGN  and around the control galaxies.   On scales larger
than a few  Mpc, AGN have almost the same  clustering amplitude as the
control sample.  This demonstrates that AGN host galaxies and inactive
control  galaxies populate  dark  matter halos  of  similar mass.   On
scales between 100  kpc and 1 Mpc, AGN are  clustered more weakly than
the  control  galaxies.   We  use  mock  catalogues  constructed  from
high-resolution  N-body  simulations   to  interpret  this  anti-bias,
showing  that the  observed effect  is  easily understood  if AGN  are
preferentially located at the centres  of their dark matter halos.  On
scales less than 70 kpc, AGN cluster marginally more strongly than the
control sample, but the effect  is weak.  When compared to the control
sample, we find that only one  in a hundred AGN has an extra neighbour
within a radius of 70 kpc.  This excess increases as a function of the
accretion rate onto the black hole, but it does not rise above the few
percent  level.    Although  interactions  between   galaxies  may  be
responsible for  triggering nuclear activity  in a minority  of nearby
AGN, some other mechanism is  required to explain the activity seen in
the majority of the objects in our sample.
\end {abstract}


\section {Introduction}

A major goal  in the study of AGN has been  to understand the physical
mechanism(s)  responsible for  triggering accretion  onto  the central
supermassive black  hole and enhanced  activity in the nucleus  of the
galaxy.  From a theoretical  standpoint, N-body simulations that treat
the  hydrodynamics of  the gas  have shown  that  interactions between
galaxies can  bring gas from  the disk to  the central regions  of the
galaxy,   leading   to   enhanced   star  formation   in   the   bulge
\citep{bh92,mh96}.  It is then natural  to speculate that some of this
gas will be accreted onto the central supermassive black hole and that
this will  trigger activity  in the nucleus  of the  galaxy.  However,
there has been little clear  observational evidence in support of this
hypothesis.

Many observational studies have  examined the correlations between AGN
activity in galaxies and  their local environment. These analyses have
produced   contradictory  results.    Early  studies   \citep[see  for
example][]{petrosian82,dahari84,keel85}  noted  that powerful  Seyfert
galaxies  appear to  show an  excess of  close companions  relative to
their  non-active counterparts.   More recent  analyses of  larger and
more complete  samples of Seyfert  galaxies have reached  the opposite
conclusion   \citep[e.g.][]{schmitt01,miller03}.   Studies   of  X-ray
selected  AGN at  intermediate  redshifts also  find  no evidence  for
excess near-neighbour  counts or  enhanced levels of  galaxy asymmetry
\citep{grogin05,waskett05}.  On the other  hand, a recent study of the
local environment  of a  sample of  $\sim 2000$ quasars  at $z  < 0.4$
drawn  from the  Sloan Digital  Sky Survey  \citep{serber06} concluded
that quasars  do have  a significant local  excess of  neighbours when
compared  to $L_*$  galaxies ,  but  only on  small scales  ($ <  0.2$
Mpc). These  authors found the  the excess to be  significantly larger
for  the   most  luminous  systems.  This  study   suggests  that  the
disagreement between different studies  may reflect the fact that they
targeted AGN with different intrinsic luminosities.

There have  also been  many studies of  the large-scale  clustering of
AGN.   This  is usually  quantified  using  the two-point  correlation
function (2PCF).   In the standard model for  structure formation, the
amplitude of the two-point  correlation function on scales larger than
a few  Mpc provides a  direct measure of  the mass of the  dark matter
halos  that host  the AGN.   The large  redshift surveys  assembled in
recent years,  in particular the two-degree Field  QSO Redshift Survey
\citep[2QZ;][]{croom01}   and    the   Sloan   Digital    Sky   Survey
\citep[SDSS;][]{york00}, have allowed the  clustering of quasars to be
studied  with unprecedented  accuracy.  The  cross-correlation between
QSOs in the  2QZ and galaxies in the  two-degree Field Galaxy Redshift
Survey  \citep[2dFGRS;][]{colless01}, measured by  \citet{croom05}, is
found to be identical to the autocorrelation of $L_*$ galaxies (a mean
bias $b_{QG}=0.97\pm0.05$). Measurements  of the two-point correlation
function  of narrow-line  AGN in  the SDSS  have been  carried  out by
\citet{wake04}. The  results are similar  to those for quasars  -- the
amplitude of  the AGN autocorrelation function is  consistent with the
autocorrelation function  of luminous galaxies  on scales from  0.2 to
100 $h^{-1}$  Mpc.  Similar results  are found for X-ray  selected AGN
\citep{gilli05,mullis05}.  On the other hand, radio-loud AGN appear to
be     significantly     more     clustered    on     large     scales
\citep{magliocchetti99,magliocchetti04,overzier03}, demonstrating that
they reside in massive dark matter halos.  This is consistent with the
fact  that  in the  local  Universe,  radio-loud  AGN are  located  in
significantly   more   massive   host   galaxies  than   optical   AGN
\citep[e.g.][]{best05}.   \citet{cv06} analyzed AGN  in the  SDSS Data
Release 2 sample and find that LINERs are more strongly clustered than
Seyfert galaxies.  Once again  this is consistent  with the  fact that
LINERs  are  found  in   more  massive  host  galaxies  that  Seyferts
\citep{kauffmann03,kewley06}.

In  this  paper,  we  analyze  the  clustering  properties  of  89,211
narrow-line AGN  selected from the Data  Release 4 (DR4)  of the Sloan
Digital    Sky    Survey   using    the    procedure   described    in
\citet{kauffmann03}.    Our  methodology  for   computing  correlation
functions in  the SDSS has  been described in detail  in \citet{li06},
where  the dependence  of clustering  on physical  properties  such as
stellar mass, age of the stellar population, concentration and stellar
surface  mass density  was studied.   In  this paper,  we extend  this
analysis to  AGN.  Our approach  differs from previous studies  in the
following ways:

\begin {enumerate}
\item We  compute AGN-galaxy cross-correlations  from scales of  a few
tens of kpc  to scales of $\sim  $10 Mpc. This allows us  to study the
detailed scale dependence of the AGN clustering amplitude.

\item We study how the clustering  depends on both the black hole mass
(estimated using  the central velocity  dispersion of the  galaxy) and
the  accretion rate relative  to the  Eddington rate  (estimated using
L[OIII], where L[OIII] is the [OIII]$\lambda$5007 line luminosity, and
$M_{BH}$ is  the black  hole mass estimated  from the  central stellar
velocity dispersion of the host).

\item We  wish to isolate the  effect of the accreting  black hole, so
the clustering is always compared  to the results obtained for control
samples of inactive galaxies that  are very closely matched to our AGN
sample.   We match  simultaneously in  redshift, mass,  and structural
properties and we test the  effect of additionally matching the age of
the stellar population  as characterized by the 4000  \AA\ break index
D$_n$(4000).

\item We  have constructed mock  catalogues using the  high resolution
Millennium  Run simulation  \citep{springel05}.   The mock  catalogues
match  the geometry  and selection  function  of galaxies  in the  DR4
large-scale structure  sample and  they also reproduce  the luminosity
and stellar mass  function of SDSS galaxies, as well  as the shape and
amplitude of the correlation functions in different bins of luminosity
and stellar mass. We use these catalogues to explore in detail how AGN
trace the underlying galaxy and halo populations.
\end {enumerate}

Our paper is  organized as follows: In section 2,  we describe the AGN
and control samples that were used  in the analysis.  In section 3, we
describe  how  we  construct   mock  catalogues  from  the  Millennium
Simulation  and  use  these  to  correct for  effects  such  as  fibre
collisions. Section  4 describes  our clustering estimator;  section 5
describes the observational results and  in section 6, we describe how
we use  our mock catalogues  to extract physical information  from the
data. Finally in  section 7, we summarize our  results and present our
conclusions.

\section { Description of Samples}

\subsection {The SDSS Spectroscopic Sample}

The data analyzed  in this study are drawn from  the Sloan Digital Sky
Survey (SDSS).  The survey goals are to obtain photometry of a quarter
of  the sky and  spectra of  nearly one  million objects.   Imaging is
obtained    in     the    {\em    u,    g,    r,     i,    z}    bands
\citep{fukugita96,smith02,ivezic04} with a  special purpose drift scan
camera  \citep{gunn98}   mounted  on  the   SDSS  2.5~meter  telescope
\citep{gunn06}  at Apache  Point  Observatory.  The  imaging data  are
photometrically     \citep{hogg01,tucker05}     and    astrometrically
\citep{pier03}  calibrated, and  used to  select stars,  galaxies, and
quasars  for follow-up fibre  spectroscopy.  Spectroscopic  fibres are
assigned to  objects on  the sky using  an efficient  tiling algorithm
designed to  optimize completeness \citep{blanton03}.   The details of
the survey strategy can be  found in \citet{york00} and an overview of
the data pipelines and products  is provided in the Early Data Release
paper \citep{stoughton02}. More details on the photmetric pipeline can
be found in \citet{lupton01}.

Our parent sample for this  study is composed of 397,344 objects which
have  been  spectroscopically  confirmed  as galaxies  and  have  data
publicly available in the SDSS Data Release~4 \citep{sdss-dr4}.  These
galaxies  are part of  the SDSS  `main' galaxy  sample used  for large
scale  structure  studies  \citep{strauss02}  and have  Petrosian  $r$
magnitudes  in the  range  $14.5 <  r  < 17.77$  after correction  for
foreground   galactic   extinction  using   the   reddening  maps   of
\citet{sfd98}.  Their  redshift distribution extends  from $\sim0.005$
to 0.30, with a median $z$ of 0.10.

The SDSS spectra are obtained with two 320-fibre spectrographs mounted
on  the SDSS  2.5-meter telescope.   Fibers 3  arcsec in  diameter are
manually plugged  into custom-drilled  aluminum plates mounted  at the
focal plane of  the telescope. The spectra are  exposed for 45 minutes
or until a fiducial signal-to-noise  (S/N) is reached.  The median S/N
per pixel  for galaxies in the  main sample is  $\sim14$.  The spectra
are  processed by  an automated  pipeline, which  flux  and wavelength
calibrates  the   data  from  3800  to   9200~\AA.   The  instrumental
resolution  is   R~$\equiv  \lambda/\delta\lambda$  =   1850  --  2200
(FWHM$\sim2.4$~\AA\ at 5000~\AA).

\subsection {The AGN and control samples}

We have performed a careful subtraction of the stellar absorption-line
spectrum  before  measuring   the  nebular  emission-lines.   This  is
accomplished by fitting the emission-line-free regions of the spectrum
with  a  model  galaxy  spectrum  computed using  the  new  population
synthesis code  of \citet{bc03}, which incorporates  a high resolution
(3 \AA\ FWHM) stellar library. A set of 39 model template spectra were
used spanning a  wide range in age and  metallicity.  After convolving
the template spectra to the measured stellar velocity dispersion of an
individual  SDSS  galaxy, the  best  fit  to  the galaxy  spectrum  is
constructed  from a  non-negative linear  combination of  the template
spectra.  Further  details are given  in \citet{tremonti04}.  Physical
parameters such  as stellar masses, metallicities,  and star formation
rates have been  estimated using the spectra and  these are publically
available  at  http://www.mpa-garching.mpg.de/SDSS/.   The  reader  is
referred  to  \citet{tremonti04}  and  \citet{brinchmann04}  for  more
details.

AGN are  selected from the  subset of galaxies  with $S/N > 3$  in the
four emission lines [OIII]$\lambda$5007, H$\beta$, [NII]$\lambda$6583,
H$\alpha$.  Following  \citet{kauffmann03}, a galaxy is  defined to be
an AGN if
\begin{equation} \log ([\rm{OIII}]/H\beta) >  
0.61/(\log ([\rm{NII}]/H\alpha)-0.05) +1.3.
\end{equation}

We divide all  the AGN into three subsamples  according to logarithmic
stellar  velocity  dispersion   $\log_{10}\sigma_\ast$.   It  is  also
interesting to study how clustering depends on the strength of nuclear
activity  in the  galaxy. In  order to  adress this  issue,  we follow
\citet{heckman04} and  use the [O {\sc iii}]  emission line luminosity
as an  indicator of  the rate  at which matter  is accreting  onto the
central  supermassive black  hole, and  we use  the relation  given in
\citet{tremaine02}  to estimate  black  hole masses  from the  stellar
velocity dispersion  measured within the fibre aperture.   We then use
the ratio $L$[O {\sc iii}]/M$_{BH}$ as a measure of the accretion rate
relative to  the Eddington rate, to define  subsamples of ``powerful''
and  ''weak'' AGN.  (Note that  in  the current  analysis, L[OIII]  is
corrected    for    dust     extinction.)     The    AGN    in    each
$\log_{10}\sigma_\ast$ subsample are  ordered by decreasing $L$[O {\sc
iii}]/M$_{BH}$.   The top  25\% are  defined as  ``powerful''  and the
bottom  25\% are  ``weak''.   For  each AGN  sample,  we construct  20
control samples  of non-AGN  by simultaneously matching  four physical
parameters: redshift, stellar mass, concentration and stellar velocity
dispersion. We  have also constructed  control samples where  the 4000
\AA\ break  strength is matched  in addition to these  parameters. The
matching tolerances  are $\Delta cz  < 500$ km s$^{-1}$,  $\Delta \log
M_* < 0.1$, $\Delta \sigma_* < 20$  km s$^{-1}$ , $\Delta C < 0.1$ and
$\Delta$D$_n$(4000)$< 0.05$.

We correct the [OIII] luminosities of the AGN in our sample
for dust using  the difference between the observed
H$\alpha$/H$\beta$ emission line flux ratios and the case-B recombination value
(2.86). We assumed  an attenuation law of the
form $\tau_{\lambda} \propto \lambda^{-0.7}$ \citep{cf00}
This procedure has clear physical meaning in the
``pure'' Seyfert 2's and LINERs. In the case of the transition
objects, the lines will arise both in the NLR and the surrounding HII
regions, with a greater relative AGN contribution to [OIII] than to the Balmer
lines. Thus,
a dust correction to [OIII] based on the ratio H$\alpha$/H$\beta$
should  be regarded as at best approximate.

\subsection{Reference galaxy sample}

We use the New York University Value Added Galaxy Catalogue (NYU-VAGC)
\footnote{http://wassup.physics.nyu.edu/vagc/}    to    construct    a
reference sample of galaxies,  which are cross-correlated with the AGN
sample.   The  original NYU-VAGC  is  a  catalogue  of local  galaxies
(mostly below $z\approx0.3$) constructed by \citet{blanton05} based on
the SDSS DR2.  Here we use a new version of  the NYU-VAGC ({\tt Sample
dr4}), which is based on SDSS DR4. The NYU-VAGC is described in detail
in \citet{blanton05}.

We have  constructed two reference  samples: 1) a  {\em spectroscopic}
reference sample ,  which is used to compute  the projected AGN-galaxy
cross-correlation function $w(r_p)$;  2) a {\em photometric} reference
sample, which is  used to calculate counts of  close neighbours around
AGN

The spectroscopic  reference sample  is constructed by  selecting from
{\tt  Sample dr4}  all  galaxies with  $14.5 <  r  < 17.6  $ that  are
identified  as  galaxies from  the  Main  sample  (note that  $r$-band
magnitude has been corrected for foreground extinction).  The galaxies
are also restricted to the redshift range $0.01\leq z\leq0.3$, and the
absolute  magnitude  range  $-23<M_{^{0.1}r}<-17$.  The  spectroscopic
reference  sample  contains  292,782  galaxies.  We  do  not  consider
galaxies fainter than $M_{^{0.1}r}=-17$, because the volume covered by
such faint samples is very small  and the results are subject to large
errors as  a result of cosmic  variance \citep[see for  example Fig. 6
of][]{li06}.  The faint apparent magnitude  limit of 17.6 is chosen to
yield uniform galaxy  sample that is complete over  the entire area of
the survey.

The photometric reference sample  is also constructed from {\tt Sample
dr4} by selecting all galaxies with $14.5<r<19$.  The resulting sample
includes 1,065,183  galaxies.  Throughout this work  we adopt standard
$\lambda$CDM     cosmological    parameters:    $\Omega     =    0.3$,
$\Omega_{\Lambda} = 0.7$, and H$_0$ = 70 km~s$^{-1}$ Mpc$^{-1}$.

\section {Mock Catalogues}

In  this section  we describe  how we  construct a  large set  of mock
galaxy samples  with the same  geometry and selection function  as the
spectroscopic samples described in the  previous section . We will use
these mock catalogues to test our method for correcting for the effect
of   fibre   collisions  on   the   measurement   of  the   AGN-galaxy
cross-correlation function  on small scales.   We will also  use these
mock catalogues  to construct models of AGN  clustering for comparison
with the observations.

\subsection{Galaxy properties in  the Millennium Simulation }

Our mock  catalogues are  constructed using the  Millennium Simulation
\citep{springel05},  a  very   large  simulation  of  the  concordance
$\Lambda$CDM   cosmogony  with   $10^{10}$   particles.   The   chosen
simulation volume is  a periodic box of size  $L_{box}=500 h^{-1}$ Mpc
on a side,  which implies a particle mass  of $8.6\times10^8$ 4$h^{-1}
M_{\odot}$.   Haloes  and  subhaloes   at  all  output  snapshots  are
identified   using   the   {\tt   subfind}  algorithm   described   in
\citet{springel01} and merger trees are then constructed that describe
how haloes grow as  the Universe evolves. \citet{croton06} implemented
a  model of  the baryonic  physics in  these simulations  in  order to
simulate  the formation and  evolution of  galaxies and  their central
supermassive  black  holes  \citep[see][for  more  details]{croton06}.
This model produced a catalogue of 9 million galaxies at $z=0$ down to
a  limiting  absolute magnitude  limit  of  $M_r-5\log h=-16.6$.  This
catalogue is well-matched to many properties of the present-day galaxy
population (luminosity-colour  distributions, clustering etc.).  It is
publically                         available                        at
http://www.mpa-garching.mpg.de/galform/agnpaper.   In   our  work,  we
adopt the positions and velocities of the galaxies given in the Croton
et al.  catalogue.  The $r$-band  luminosities and stellar  masses are
assigned  to  each  model  galaxy, using  the  parametrized  functions
described  by  \citet{wang06}.  These  functions relate  the  physical
properties of  galaxies to the  quantity $M_{infall}$, defined  as the
mass of  the halo at  the epoch when  the galaxy was last  the central
dominant object in its own halo.  They were chosen so as to give close
fits  to  the  results  of  the physical  galaxy  formation  model  of
\citet{croton06}, but their coefficients  were adjusted to improve the
fit to the SDSS data, in particular the galaxy mass function at the low
mass end.   Extensive tests have  shown that the  adopted parametrized
relations allow us to accurately match the luminosity and stellar mass
functions of galaxies in the SDSS,  as well as the shape and amplitude
of  the  two  point  correlation  function of  galaxies  in  different
luminosity and stellar mass ranges \citep{li06,wang06}.

\subsection{Constructing the catalogues}                                    

Our aim  is to  construct mock galaxy  redshift surveys that  have the
same  geometry and  selection function  as the  SDSS DR4.   A detailed
account  of  the   observational  selection  effects  accompanies  the
NYU-VAGC  release.  The  survey  geometry  is expressed  as  a set  of
disjoint convex spherical polygons, defined by a set of ``caps''. This
methodology was  developed by Andrew  Hamilton to deal  accurately and
efficiently  with   the  complex  angular  masks   of  galaxy  surveys
\citep{ht02}
\footnote{http://casa.colorado.edu/$^\sim$ajsh/mangle/}.            The
advantage of using this method is that it is easy to determine whether
a point is  inside or outside a given  polygon \citep{tegmark02}.  The
redshift  sampling  completeness is  then  defined  as  the number  of
galaxies with  redshifts divided by the total  number of spectroscopic
targets  in the  polygon.  The  completeness is  thus  a dimensionless
number  between  0 and  1,  and  it is  constant  within  each of  the
polygons.  The limiting magnitude in each polygon is also provided (it
changes slightly across the survey region).

We  construct  our mock  catalogues  using  the  methods described  in
\citet{yang04}, except that we  position the virtual observer randomly
inside the  simulation and not at  the centre of the  box. Because the
survey extends out to $z \sim 0.3$, this implies that we need to cover
a volume that extends to a  depth of $900 h^{-1}$ Mpc, i.e. twice that
of  the Millennium  catalogue.   We thus  create  $5\times 5\times  5$
periodic  replications of the  simulation box  and place  the observer
randomly within  the central  box, so that  the required depth  can be
achieved in all directions for the observer.

We  produce 20 mock  catalogues by  following the  procedure described
below:
\begin{enumerate}
\item  We randomly  place a  virtual observer  in the  stack  of boxes
described above. We  define a ($\alpha$,$\delta$)-coordinate frame and
remove all galaxies that lie outside the survey region.
\item For each galaxy we compute the redshift as "seen" by the virtual
observer.  The redshift is determined by the comoving distance and the
peculiar velocity of the galaxy.
\item We compute  the $r$-band apparent magnitude of  each galaxy from
its absolute  magnitude $M_r$ and its redshift,  applying a (negative)
K-correction  but  neglecting  any  evolutionary correction.  We  then
select  galaxies according to  the position-dependent  magnitude limit
(provided in the {\tt Sample dr4}) and apply a (positive) K-correction
to  compute  $M_{^{0.1}r}$, the  $r$-band  absolute  magnitude of  the
galaxy at $z=0.1$.
\item  To  mimic  the  position-dependent  completeness,  we  randomly
eliminate  galaxies  using the  completeness  masks  provided in  {\tt
Sample dr4}.
\item  Finally, we  mimic the  actual  selection criteria  of our  own
reference  sample by  restricting galaxies  in the  mock  catalogue to
$0.01<z<0.3$, $14.5<r<17.6$ and $-23< M_{^{0.1}r}<-17$.
\end{enumerate}

\begin{figure*}
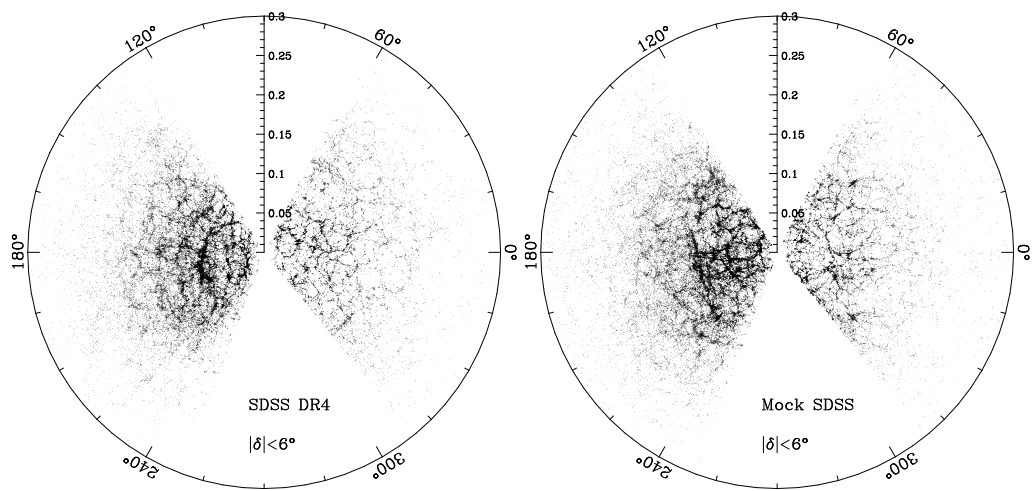

\centerline{
\psfig{figure=f1a.ps,width=0.8\hssize}
\psfig{figure=f1b.ps,width=0.8\hssize}}
\caption{Equatorial distribution  of right ascension  and redshift for
galaxies within 6$^\circ$ of the equator in the SDSS (left) and in one
of our mock catalogues (right).}
\label{fig:mock}
\end{figure*}

Figure~\ref{fig:mock} shows the equatorial distribution of galaxies in
one  of our  mock catalogues,  compared to  that in  the observational
sample.   The average  number of  galaxies in  our mock  catalogues is
$\sim$320,000,  with  a  r.m.s.   dispersion of  $\sim9000$,  in  good
agreement with the observed number.

\subsection{Fibre collisions}

The procedure described  above does not account for  the fact that the
spectroscopic  target  selection  becomes increasingly  incomplete  in
regions  of the  sky where  the galaxy  density is  high,  because two
fibres cannot be positioned closer than 55 arcseconds from each other.
In  order to  mimic these  fibre  ``collisions'', we  modify step  (4)
above.  We  no longer randomly sample galaxies  using the completeness
masks.   Instead,  we assign  fibres  to  our  mock galaxies  using  a
procedure  that is  designed to  mimic  the tiling  code that  assigns
spectroscopic fibres to SDSS target galaxies \citep{blanton03}. We run
a friends-of-friends  grouping algorithm on  the mock galaxies  with a
55$^{\prime\prime}$ linking  length. Isolated galaxies,  i.e. those in
single-member groups, are are  always assigned fibres.  If two targets
collide,  we simply  pick one  at random  to be  the spectroscopically
targeted galaxy.   For groups with more  than two members,  we need to
determine which  members will be  assigned fibres and which  will not.
The   resulting  groups   are  almost   always  of   sufficiently  low
multiplicity  so  that  in  principle,  one  could  simply  check  all
possibilities to  find the best possible combination  of galaxies that
would eliminate fibre collisions (this is the procedure adopted in the
SDSS target selection).  In our  work we use a somewhat more efficient
method. For each group, we  calculate the geometric centre.  The group
members located  closer to  the center are  preferentially eliminated.
For example, in a triple  collision this algorithm will keep the outer
two members rather than the middle one.

This procedure  is complicated by the  fact that some  fraction of the
sky  will   be  covered  with   overlaps  of  different   tiles  (Each
spectroscopic fibre plug plate is  referred as a ``tile'', which has a
circular field of  view with a radius of  $1^\circ_.49$.). About 30 \%
of  the sky  is covered  in such  overlaps.  This  means that  if, for
example, a  binary group covered  by two ore  more tiles, both  of the
group members  can be  assigned fibres. We  take this into  account by
iteratively repeating the procedure described above, as follows.
\begin{enumerate}
\item  First, we  determine the  number of  tiles that  cover  a given
galaxy and  set the quantity  $n_{chances}$ equal to this  number. For
example,  $n_{chances}=2$ for  a  galaxy covered  by  two tiles;  this
galaxy has two chances to be assigned a fibre.
\item We  assign fibres by  applying the algorithm described  above to
those galaxies  with $n_{chances}>0$. If  a galaxy obtains a  fibre in
this procedure,  then the quantity  $n_{chances}$ is set to  zero.  If
not, we set $n_{chances}=n_{chances}-1$.
\item Step (ii)  is repeated until $n_{chances}$ reaches  zero for all
galaxies. All galaxies  that are not assigned fibres  are then removed
from our mock catalogue.
\item Finally,  we remove a number  of galaxies at random  so that the
resulting  mock   sample  has  the   same  overall  position-dependent
completeness as the real SDSS sample.
\end{enumerate}

\section{Clustering measures}

In order  to compute  the two-point cross-correlation  function (2PCF)
$\xi(r_p,\pi)$ between  the AGN host  (or matched control)  sample and
the reference galaxy sample ,  we have constructed random samples that
are designed to include  all observational selection effects.  This is
described   in  detail  in   \citet{li06}.   $\xi(r_p,\pi)$   is  then
calculated using the estimator
\begin{equation}
\xi(r_p,\pi) = \frac{N_R}{N_D} \frac{QD(r_p,\pi)}{QR(r_p,\pi)} -1,
\end{equation}
where $r_p$  and $\pi$ are the separations  perpendicular and parallel
to the  line of sight; $N_D$ and  $N_R$ are the number  of galaxies in
the  reference sample  and  in the  random  sample; $QD(r_p,\pi)$  and
$QR(r_p,\pi)$ are the  cross pair counts between AGN  (or control) and
the  reference sample,  and between  AGN (or  control) and  the random
sample, respectively.

In what  follows, we focus  on the projection of  $\xi(r_p,\pi)$ along
the line of sight:
\begin{equation}
w_p(r_p)=\int_{-\infty}^{+\infty}\xi(r_p,\pi)d\pi=
\sum_i\xi(r_p,\pi_i)\Delta\pi_i.
\end{equation}
Here the summation for computing $w_p(r_p)$ runs from $\pi_1 = -39.5 $
h$^{-1}$ Mpc to $\pi_{80} = 39.5$ h$^{-1}$ Mpc, with $\Delta\pi_i = 1$
h$^{-1}$ Mpc.

The  errors on  the clustering  measurements are  estimated  using the
bootstrap  resampling   technique  \citep{bbs84}.   We   generate  100
bootstrap samples  from the  observations and compute  the correlation
functions  for each  sample using  the weighting  scheme (but  not the
approximate  formula) given  by  \citet{mjb92}.  The  errors are  then
given  by  the  scatter  of  the measurements  among  these  bootstrap
samples.   More  details  about   our  procedures  and  tests  of  our
methodology  can  be found  in  \citet{li06}.  We  also obtain  robust
estimates of  our uncertainties from  the scatter between  the results
obtained from disjoint areas of the sky.

\begin{figure}
\centerline{\psfig{figure=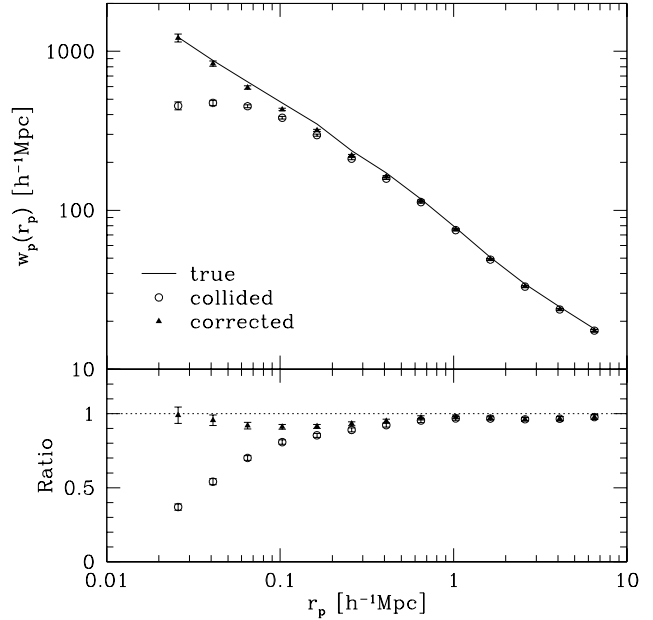,width=\hssize}}
\caption{Top: Projected  2PCF $w_p(r_p)$  measured for  the  mock catalogue
without  including fibre  collisions  (solid line)  and  for the  mock
catalogue with  fibre collisions (circles). The  filled triangles show
the measured  $w_p(r_p)$ for the  mock after correcting the  effect of
fibre collisions using the method described in the text.
{Bottom: The ratios of the uncorrected and the corrected $w_p(r_p)$ relative
to the "true" $w_p(r_p)$.}}
\label{fig:wrp_mock}
\end{figure}

We are  particularly interested in the amplitude  of the AGN-reference
galaxy cross-correlation  on small  scales ($<100$ kpc),  because this
allows us to evaluate whether  mergers and interactions play a role in
triggering AGN activity.  A careful correction for the effect of fibre
collisions when  measuring the clustering  is thus very  important. As
described  in  \citet{li06},  we   correct  for  fibre  collisions  by
comparing the  angular 2PCF of  the spectroscopic sample with  that of
the parent photometric sample. Here we use our mock SDSS catalogues to
test  the correction  method.   We calculate  the angular  correlation
functions $w_z(\theta)$  and $w_p(\theta)$ using  mock catalogues with
and without fibre collisions. The function
\begin{equation}
F(\theta)=\frac{1+w_z(\theta)}{1+w_p(\theta)}
\end{equation}
is  then used  to correct  for collisions  by weighting  each  pair by
$1/F(\theta)$.    Figure~\ref{fig:wrp_mock} (the top panel)
shows  measurements   of
$w_p(r_p)$ for galaxies in one of our mock catalogues.  The solid line
is  the   ``true''  correlation  function  calculated   for  the  mock
catalogues  that do not  include fibre  collisions.  Circles  show the
results  when  fibre  collisions  are included.   Triangles  show  the
results that are obtained when the 2PCF is corrected for the effect of
fibre collisions  using the method  described above. 
In the bottom panel, we plot the ratios of the uncorrected and
the corrected $w_p(r_p)$ relative to the "true" correlation function.
 As can  be seen,
our correction procedure works well.  The turnover in the amplitude on
small physical scales resulting from the lower sampling of galaxies in
dense regions disappears and the corrected $w_p(r_p)$ is very close to
the real one. It is noticeable that there is still a very small
deficit in the corrected $w_p(r_p)$ on scales between 0.05 and 1 Mpc.
This should not be a significant contribution to the bias between AGN
and normal galaxies (see below), because fibre collisions are expected
to affect the AGN and the reference galaxies in the same way.

\section{Observational results}

\subsection{AGN Bias}

We  first compute $w_p^{\rm  AGN/ref}(r_p)$, the  cross-correlation of
the AGN sample with respect  to the reference sample.  As described in
section 2,  we have constructed two  sets of 20  control samples.  The
first set is constructed  by simultaneously matching redshift, stellar
mass, concentration  and stellar  velocity dispersion, and  the second
set  by additionally matching  the 4000\AA\  break strength.   We then
compute     $\bar{w}_p^{\rm      contr/ref}(r_p)$,     the     average
cross-correlation of the control samples with respect to the reference
sample.     The   quantity    $w_p^{\rm   AGN/ref}(r_p)/\bar{w}_p^{\rm
contr/ref}(r_p)$ then measures  the {\em bias} of the  AGN sample with
respect to  the control sample of  non-AGN as a  function of projected
radius $r_p$.

The results are  shown in Fig. 3. In the top  panel, we plot $w_p^{\rm
AGN/ref}(r_p)$   as  circles.   $\bar{w}_p^{\rm   contr/ref}(r_p)$  is
evaluated  for the two  sets of  control samples  and the  results are
plotted as squares for the first set and triangles for the second set.
The measurement  errors are  estimated using the  bootstrap resampling
technique described in the previous  section.  In the bottom panel, we
plot    the     ratio    $    w_p^{\rm    AGN/ref}(r_p)/\bar{w}_p^{\rm
contr/ref}(r_p)$ for the two control samples, The errors are estimated
in the  same manner  as in  the top panel.   For clarity,  squares and
triangles  in  both  panels  have  been  slightly  shifted  along  the
$r_p$-axis.

\begin{figure}
\centerline{\psfig{figure=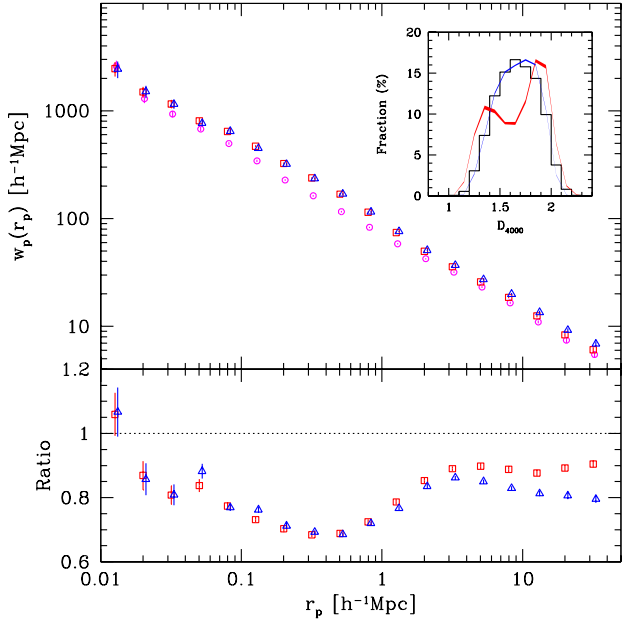,width=\hssize}}
\caption{Top:    projected   cross-correlation    function   $w_p^{\rm
AGN/ref}(r_p)$  between  AGN  and  reference galaxies  is  plotted  as
circles. The average cross-correlation  between two sets of 20 control
samples  of non-AGN  and the  same  reference galaxies  is plotted  as
squares  and  triangles  (squares:  redshift,  stellar  mass,  stellar
velocity  dispersion  and concentration  are  matched; triangles:  the
control   samples  are  constructed   by  additionally   matching  the
4000\AA-break   strength).    The   inset  compares   the   D$_{4000}$
distribution for  the two  sets of control  samples, with red  for the
former  set  and  blue  for   the  latter.  The  histogram  shows  the
D$_n$(4000) distribution for the AGN.  Bottom: ratio of the $w_p(r_p)$
measurement of AGN to that of non-AGN.  Symbols are the same as in the
top panel.}
\label{fig:all}
\end{figure}

Figure 3 shows  that there exists a {\em  scale-dependent bias} in the
distribution  of  AGN  relative   to  that  of  normal  galaxies.   In
particular, the  ratio between  the two cross-correlations  appears to
exhibit a pronounced  ``dip'' at scales between 100 kpc  and 1 Mpc. We
note that errorbars estimated using the bootstrap resampling technique
do not take into account effects due to cosmic variance. The coherence
length of the  large scale structure is large and even  in a survey as
big  as the  SDSS, this  can  induce significant  fluctuations in  the
amplitude  of the correlation  function from  one part  of the  sky to
another.   The  difference in  the  clustering  amplitude  of AGN  and
non-AGN  shown in  Figure  3  is only  a  10-30\% effect  ,  so it  is
important to test whether the dip seen in Figure 3 is truly robust.

We have  thus divided the  survey into 6  different areas on  the sky.
Each subsample includes  $\sim 12,000$ AGN. We recompute  the AGN bias
for each  of these  subsamples and  the results are  shown in  Fig. 4.
Note that in this plot, we only use a single control sample to compute
the bias, not  the average of 20 control samples as  in Figure 3.  The
scatter  between the  different curves  in Figure  4 thus  provides an
upper limit to  the true error in the measurement of  the bias. As can
be seen,  on small  scales ($<  $ 50 kpc)  , the  different subsamples
scatter in  bias above and below  unity.  On scales between  0.1 and 1
Mpc, however, all 6 subsamples  lie systematically below this line. On
scales  larger than  1-2 Mpc,  five out  of six  subsamples  show bias
values below unity, but the effect is clearly less significant than on
scales between 0.1  and 1 Mpc.  In Fig.  5,  we examine the dispersion
in  the bias  measurement for  the  AGN sample  as a  whole caused  by
differences between the  control samples. As can be  seen, the scatter
in the  measurement of the  bias between different control  samples is
considerably  smaller  than   the  scatter  between  different  survey
regions, showing that  that cosmic variance is, in  fact, the dominant
source  of uncertainty  in our  results.  Once  again, there  is clear
indication that AGN are antibiased relative to the control galaxies on
scales larger than 100 kpc.

\begin{figure}
\centerline{\psfig{figure=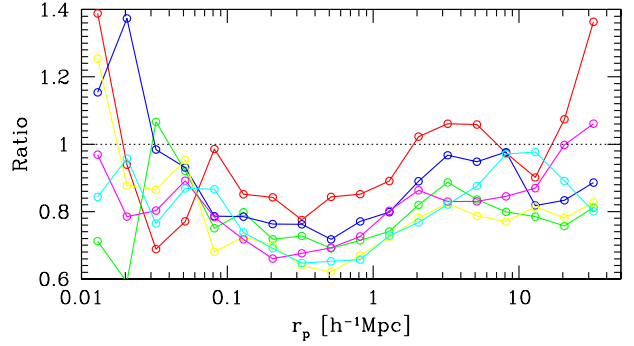,width=\hssize}}
\vspace{-3.5cm}
\caption{  Ratio of  the  $w_p(r_p)$  measurement of  AGN  to that  of
non-AGN for six disjoint regions of the survey.}
\label{fig:areas}
\end{figure}

We  conclude  that   on  scales  between  0.1  and   1  Mpc,  AGN  are
 significantly  anti-biased relative  to non-AGN  of the  same stellar
 mass, concentration and stellar  velocity dispersion.  Figure 3 shows
 that this  anti-bias persists even when  the mean age  of the stellar
 population  is matched  in addition  to stellar  mass  and structural
 parameters.   We  note  that  these  scales  are  comparable  to  the
 diameters of the dark matter halos that are expected to host galaxies
 with  stellar masses  comparable to  the objects  in our  sample.  In
 section  6, we  construct  halo occupation  (HOD)  models using  mock
 catalogues  constructed  from  the  Millennium  Simulation  that  can
 explain the  anti-bias on  these scales.  As  we will show,  the same
 models  naturally  predict a  smaller,  but  significant antibias  on
 scales larger than 1 Mpc.

\subsection{Dependence on black hole mass and AGN power }

\begin{figure*}
\centerline{\psfig{figure=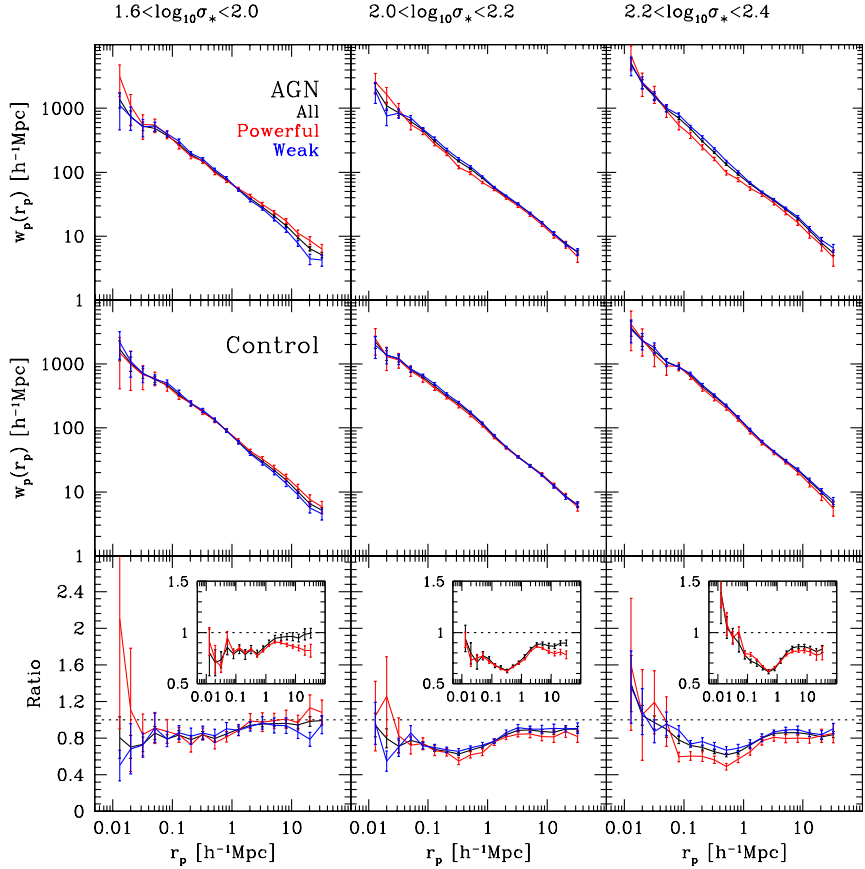,width=0.7\hdsize}}
\caption{Top:  projected  cross-correlation  $w_p(r_p)$  in  different
$\sigma_\ast$ bins (indicated above  each panel), for all AGN (black),
powerful  (red) and  weak  (blue)  AGN. The  powerful  (weak) AGN  are
defined as the top (bottom)  25 per cent objects ordered by decreasing
$L$[O {\sc iii}]$/M_\bullet$. The middle row is for control samples of
non-AGN and the bottom row shows the ratio between the results for the
AGN and the control samples.   The insets in the bottom panels compare
results for  all AGN  using different control  samples.  Black  is for
control  samples  constructed  by  matching  redshift,  stellar  mass,
stellar  velocity  dispersion  and  concentration, while  red  is  for
control samples where the 4000\AA\-break strength is also matched.}
\label{fig:sig}
\end{figure*}

\begin{figure*}
\centerline{\psfig{figure=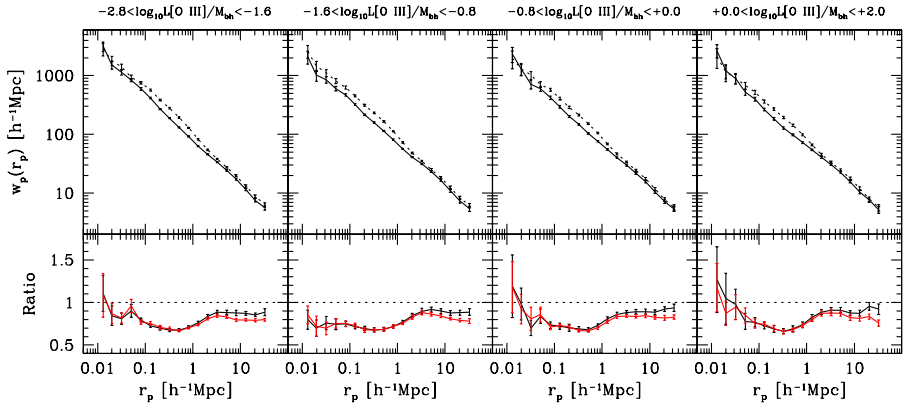,width=0.7\hdsize}}
\vspace{-6cm}
\caption{$w_p(r_p)$  in different  $L$[L  {\sc iii}]$/M_\bullet$  bins
(indicated  above each  panel), for  all AGN  (solid) and  for control
samples of non-AGN (dashed).  The  small panels give the ratio between
the  above  two  (black).   The   red  lines  are  results  where  the
4000\AA-break  strength  is  also  matched when  constructing  control
samples.}
\label{fig:edd}
\end{figure*}

It is  interesting to study how  AGN clustering depends  on black hole
mass and the strength of  nuclear activity in the galaxy. As described
above, we use  the stellar velocity dispersion as  an indicator of the
black hole mass and divide  all AGN into three subsamples according to
$\log_{10}\sigma_\ast$.    We   then   use   the  ratio   $L$[O   {\sc
iii}]/M$_{BH}$  as a  measure of  the accretion  rate relative  to the
Eddington  rate. We  rank order  all the  AGN in  a given  interval of
stellar velocity dispersion according to $L$[O {\sc iii}]/M$_{BH}$ and
we  define  subsamples  of  ``powerful''  and ''weak''  AGN  as  those
contained  within  the  upper   and  lower  25th  percentiles  of  the
distribution of this quantity.

\begin{figure}
\centerline{\psfig{figure=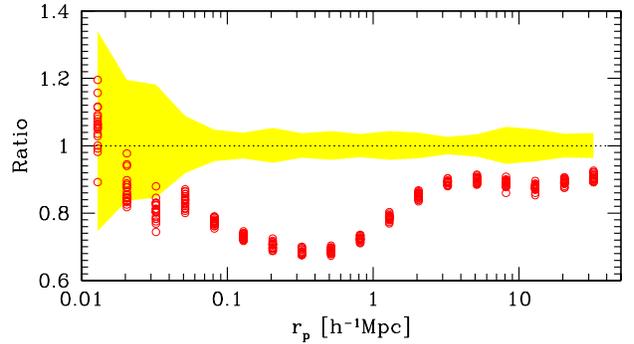,width=\hssize}}
\vspace{-3.5cm}
\caption{Ratio of the $w_p(r_p)$ measurement of AGN to that of non-AGN
for 20  different non-AGN control  samples.  The yellow  shaded region
shows  the  ratio of  the  $w_p(r_p)$  measurements  of the  different
non-AGN control samples relative to each other.}
\label{fig:ctrl_disp}
\end{figure}

The  results  are  shown in  Figure  6.   Panels  from left  to  right
correspond  to  different   intervals  of  $\log_{10}\sigma_\ast$,  as
indicated  at the  top of  the figure.   The first  two rows  show the
$w_p(r_p)$  measurements for  the  AGN and  the corresponding  control
samples. The  third row shows the  ratio between the  two.  Red (blue)
lines correspond to the  powerful (weak) subsamples.  Black lines show
results for the sample as a whole.

As mentioned previously, there are two sets of control samples: sample
1 is  constructed by  simultaneously matching redshift,  stellar mass,
concentration and stellar velocity dispersion; sample 2 is constructed
by additionally matching the 4000\AA\ break strength. For clarity, the
main panels  in in Figure 6 show  the results only for  sample 1.  The
insets   in   the   bottom   row   compare  the   ratio   $   w_p^{\rm
AGN/ref}(r_p)/\bar{w}_p^{\rm  contr/ref}(r_p)$  for  the  two  control
samples  (black corresponds  to  set 1  and  red to  set  2). The  two
different  control samples give  very similar  results on  scales less
than a  few Mpc, but the  large-scale antibias is  more pronounced for
sample 2 (note that this is also seen in Figure 2).

As can be seen, the ``dip''  in clustering on scales between 0.1 and 1
Mpc  is most  pronounced  for  AGN with  the  largest central  stellar
velocity  dispersions  and the  highest  accretion  rates.  On  scales
smaller  than 0.1  Mpc,  more  powerful AGN  appear  be somewhat  more
strongly clustered  than weaker AGN  and more strongly  clustered than
galaxies in  the control sample.   The error bars on  the measurements
are large, however, and effect is not of high significance.  In Figure
7 we plot results for AGN of all velocity dispersions, but now in four
different intervals of $L$[O  {\sc iii}]/M$_{BH}$, as indicated at the
top of the figure.  Once again we see a marginal tendency for AGN with
higher  values  of  $L$[O  {\sc  iii}]/M$_{BH}$ to  be  more  strongly
clustered on small scales.

\subsection {Close Neighbour Counts}

\begin{figure*}
\centerline{\psfig{figure=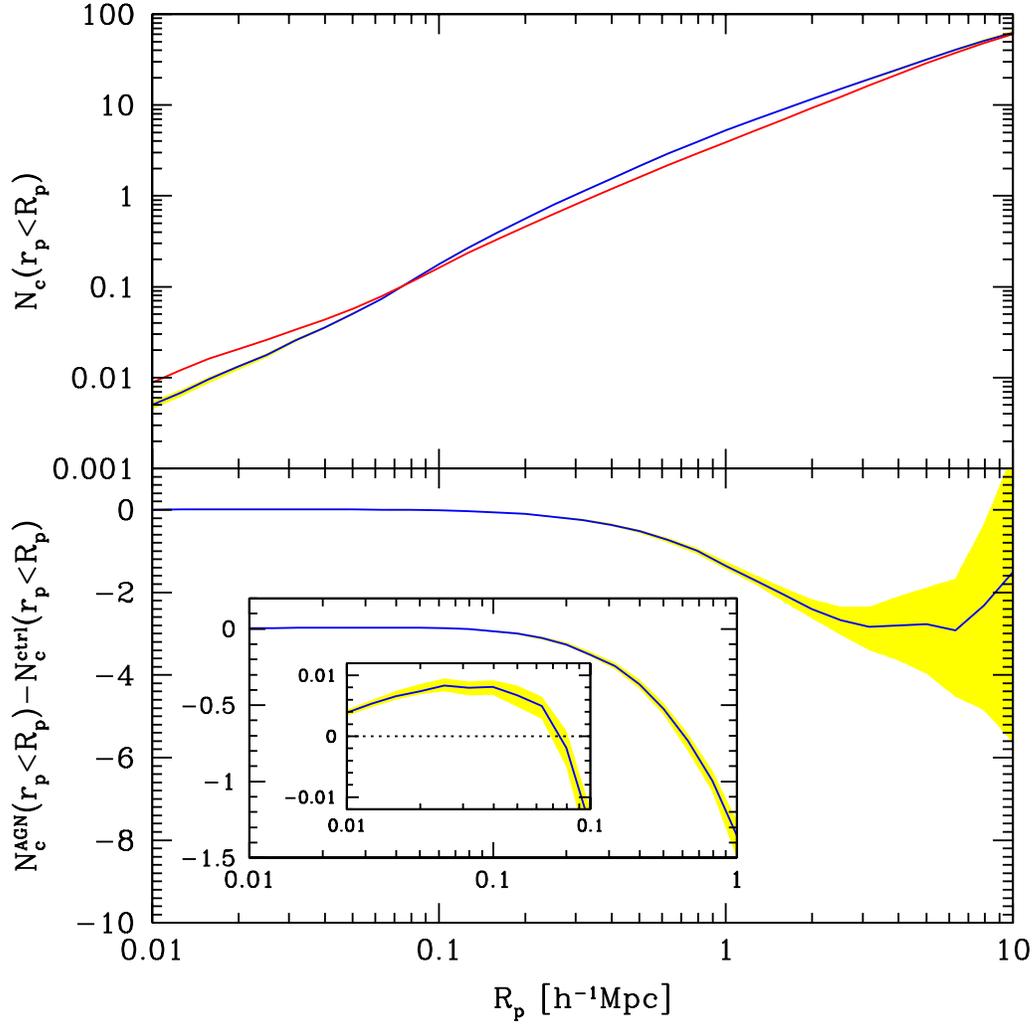,width=0.8\hdsize}}
\caption{ Top:  Average counts of  galaxies in the  photometric sample
($r_{lim} <  19$) within a given  projected radius $R_p$  from the AGN
(red)  and from the  control galaxies  (blue). Bottom:  The difference
between the counts around the  AGN and the control galaxies is plotted
as a function of $R_p$.  The yellow bands indicate the variance in the
results between the 20 different control samples.}
\label{fig:counts_all}
\end{figure*}

\begin{figure*}
\centerline{\psfig{figure=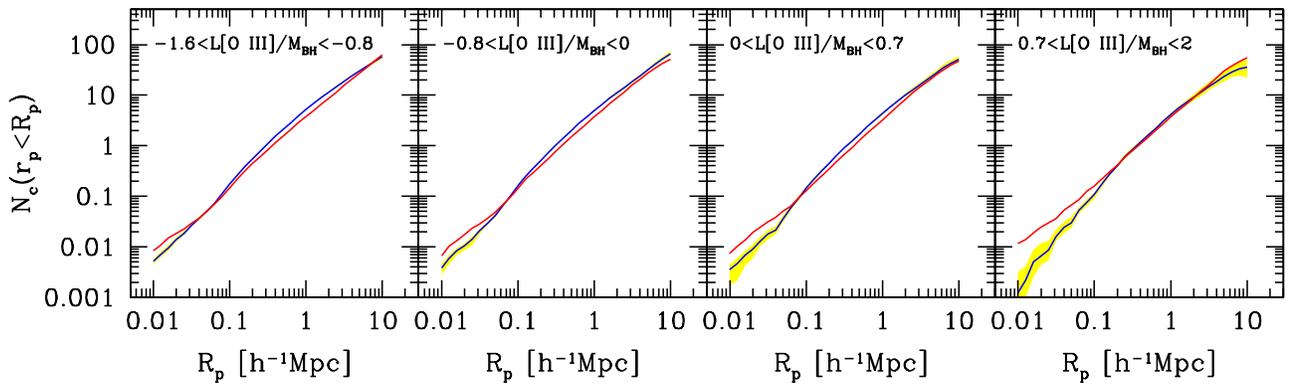,width=\hdsize}}
\vspace{-12cm}
\caption{Same as  the top panel of  Fig.~\ref{fig:counts_all}, but for
four subsamples of AGN with different $L$[O {\sc iii}]$/M_\bullet$, as
indicated in each panel.}
\label{fig:counts_edd}
\end{figure*}

As  we  have  discussed,  one  of  the  problems  with  computing  the
galaxy-AGN cross-  correlation function on very  small physical scales
in the SDSS is that corrections for the effect of fibre collisions are
required.   These corrections are  statistical in  nature and  even if
they  are correct  on  average, they  may  still introduce  systematic
effects  in our  analysis.  An  alternative approach  is to  count the
number of galaxies  in the vicinity of AGN  in the photometric sample,
which is not affected by incompleteness. The disadvantage of using the
photometic sample  is that  many of the  close neighbours will  not be
truly nearby systems, but  rather chance projections of foreground and
background galaxies that  lie along the line-of-sight.  We  can make a
statistical  correction  for  this  by evaluating  the  counts  around
randomly placed ``galaxies'' with  the same assumed joint distribution
of apparent magnitude and redshift as the AGN (control) samples.

In  the  upper  panel of  Figure  8  we  plot the  average  correlated
neighbour  count (i.e. after  statistical correction  for uncorrelated
projected  neighbours) within a  given value  of the  projected radius
$R_p$ for  the AGN  sample (red) and  the control samples  (blue). The
lower panel  and its insets  show the difference betweeen  the average
correlated counts  for the  AGN and control  samples as a  function of
$R_p$.   The  variance  in  the  counts around  the  control  galaxies
estimated from  the 20 different  control samples is shown  in yellow.
The  AGN sample  has a  $r$-band limiting  magnitude of  17.6  and the
photomteric reference sample  that we use is limited  at $r$=19.0.  In
order  to  ensure that  we  are  counting  similar neighbours  at  all
redshifts, the  counts only  include those galaxies  with $r  < r_{\rm
AGN} +1.4$  mag.  In this analysis,  the control sample  is matched in
$r$-band {\em  apparent} magnitude as well as  redshift, stellar mass,
velocity  dispersion  and concentration.   This  ensures  that we  are
counting galaxies to  the same limiting magnitude around  both the AGN
and the control galaxies.

Figure 8 shows that the counts around the AGN and the control galaxies
match well on large scales.  On small scales, there is  a small but 
significant excess in the number of neighbours around AGN out to 
scales of  $\sim 70$ kpc.  As may be seen from the bottom panel of 
Figure 8, 
AGN are appromimately  twice as likely to have a near neighbour
as galaxies in the control sample. This does not mean, however,
that every AGN has a close companion. Figure 8 also shows that only one in a
hundred AGN has  an additional  close ($R_p < 70$ kpc)   neighbour as  
compared to  the control  galaxies. On scales larger than 100 kpc, the 
pair counts around the AGN dip below  the counts around the control 
samples, leading to the ``anti-bias'' discussed in the previous 
section.  This may be compensated on scales larger than several Mpc, 
although such compensation is not required with our present statistics.

Figure 9 shows the counts around AGN in four different ranges of
L[OIII]/$M_{BH}$. As can be seen, the excess on small scales increases
as a function of the accretion rate onto the black hole. However, the 
excess affects only a few percent of the AGN, even for the objects in 
our highest L[OIII]/M$_{BH}$ bin.  We note that \citet{serber06} 
analyzed galaxy counts around quasars compared to $L_*$ galaxies at 
the same redshift  and found a clear excess on scales less than 100 
kpc, very  similar to the $\sim 70$ kpc scales where we see the upturn 
in the counts around our sample of narrow-line AGN.  Serber et 
al also found that the excess was largest for the most luminous 
quasars; the excess count reached values $\sim 1$ (i.e. significantly 
larger than the excess found for the most powerful narrow-line AGN in 
our sample) for quasars with $i$-band magnitudes brighter than $-24$.
If we use the relation between [OIII] line luminosity and quasar 
continuum luminosity of \citet{zakamska03} to compare the AGN in our 
sample to the quasars studied by Serber et al, we find that the 
luminosities where quasars begin to exhibit a significant excess count 
lie just beyond those of the AGN that populate our highest
L[OIII]/M$_{BH}$ bin.

Our conclusion, therefore, is that 
we do not find strong evidence that interactions and mergers
are  playing a significant
role in triggering  the activity  in typical  AGN in the local Universe.
One caveat that should be mentioned is that if the activity is
triggered {\em after} the merger has already taken place,
our pair count statistics would not be a good diagnostic. In order to
assess this, more work is needed to assess whether AGN exhibit any evidence
for disturbed morphologies or stuctural peculiarities.

\section {Interpretion of AGN Clustering using Halo Occupation Models}

In  the previous section  we showed  that the  main difference  in the
AGN-galaxy  cross-correlation  function  with  respect to  that  of  a
closely matched control sample of  non-AGN is that AGN are more weakly
clustered on scales between 100 kpc and 1 Mpc. On larger scales, there
is  a much  smaller difference  in the  clustering signal  of  AGN and
non-AGN. The  clustering amplitude of  AGN on large scales  provides a
measure  of  the  mass  of  the  dark matter  halos  that  host  these
objects. The  fact that there is  only a small  difference between the
AGN and  the control  sample tells  us that AGN  are found  in roughly
similar dark matter halos to  non-AGN with the same stellar masses and
structural properties.

\begin{figure}
\centerline{\psfig{figure=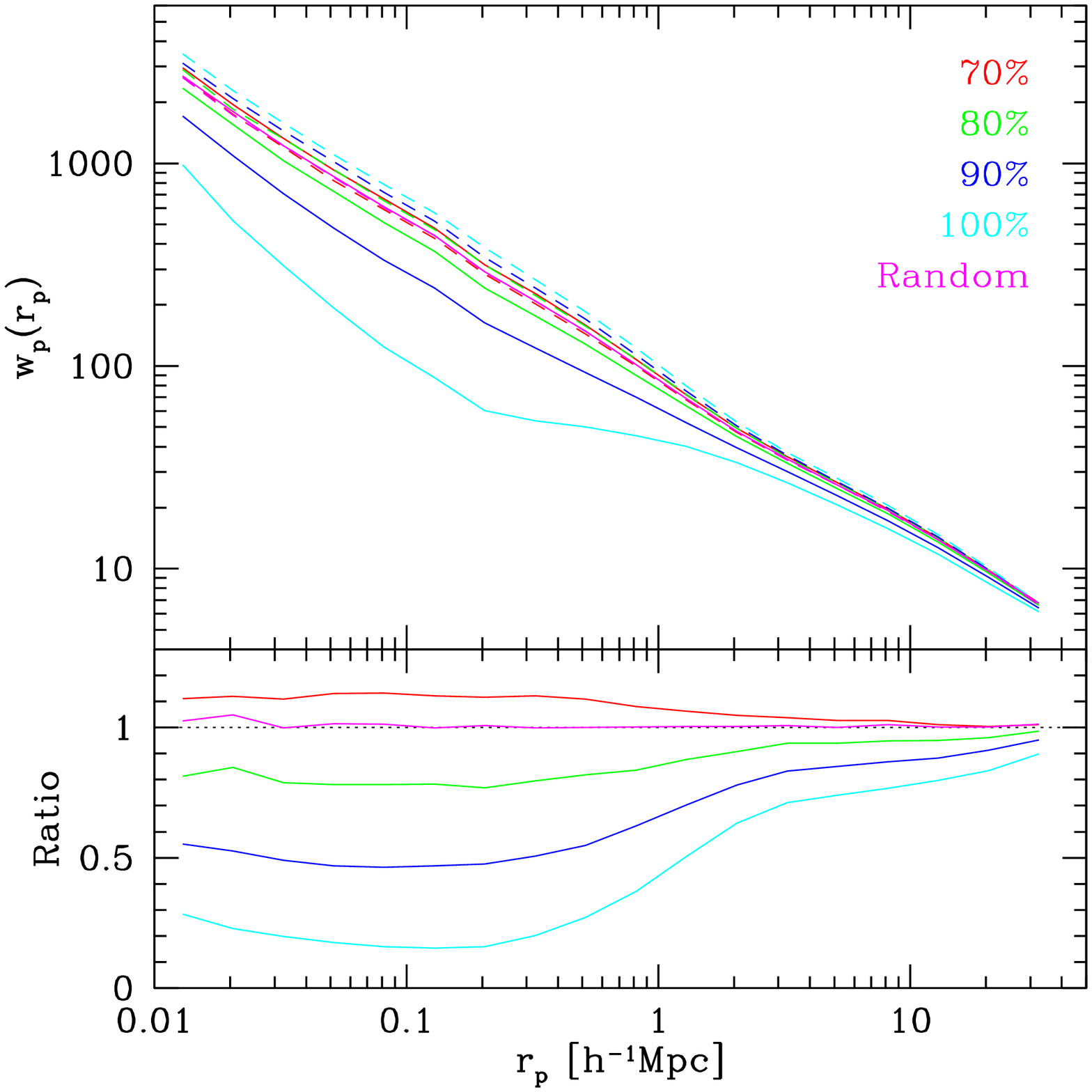,width=\hssize}}
\caption{ Top: The AGN/reference cross-correlation function calculated
from  the  mock catalogues  is  plotted  as  solid curves,  while  the
control/reference  cross-correlations are  plotted  as dashed  curves.
The different colours indicate models  in which a given percentage (as
indicated on the plot) of the  AGN are located at the centers of their
own dark  matter halos.  Bottom:  The ratio between  the AGN/reference
galaxy cross-correlation  functions and the  control galaxy/ reference
galaxy correlation functions are plotted for the same set of models.}
\label{fig:model}
\end{figure}

\begin{figure}
\centerline{\psfig{figure=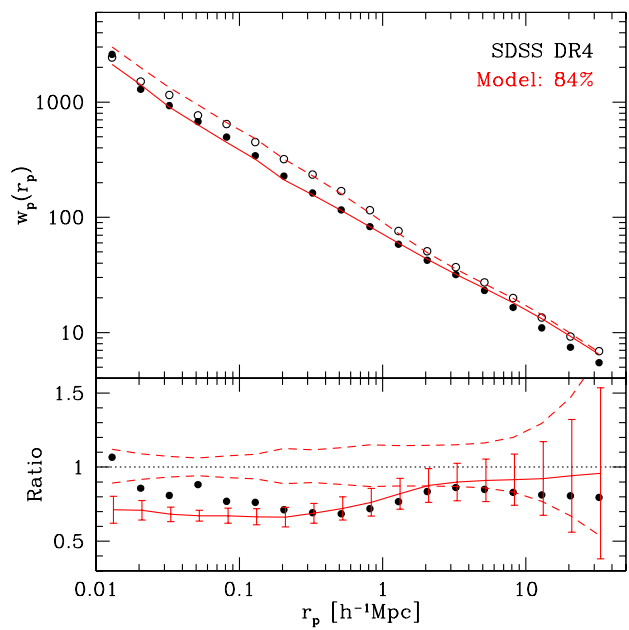,width=\hssize}}
\caption{ Top: The AGN/reference cross-correlation function calculated
from the SDSS  is plotted as solid circles. The  open circles show the
control  galaxy/reference galaxy  cross-correlation function  from the
SDSS.  Results for AGN and control galaxies for our best-fit model are
plotted as solid and dashed red lines.  Bottom: Solid circles show the
ratio  $w^{\rm AGN/ref}(r_p)/w^{\rm  control/ref}(r_p)$  for the  SDSS
sample.   The  solid red  curve  shows  the  result for  our  best-fit
models. The error bars indicate the uncertainty due to cosmic variance
as  estimated from  20 different  mock catalogues  (see text  for more
details).   The  dashed  red  curves  indicate  the  variance  between
different control samples from different mock catalogues (see text).}
\label{fig:model_obs}
\end{figure}

The physical scales of 0.1-1 Mpc where we do see strong differences in
the clustering of AGN and  non-AGN are similar to the virial diameters
of  the dark  matter halos  that are  expected to  host  galaxies with
luminosities  of   $\sim  L_*$  \citep{mandelbaum06}.    The  simplest
interpretation  of the AGN  anti-bias on  these scales,  therefore, is
that  AGN occupy preferred  positions within  their dark  matter halos
where  conditions are more  favourable for  continued fuelling  of the
central black hole.  One obvious  preferred location would be the halo
centre  where  gas  is  expected  to  be able  to  reach  high  enough
overdensities to  cool via radiative  processes.  In the main  body of
the more massive  dark matter halos, most of  the surrounding gas will
have been shock heated to the  virial temperature of the halo and will
no longer be able to cool efficiently.  In addition, the vast majority
of galaxy-galaxy mergers within a halo will occur with the galaxy that
is located at the halo centre \citep{springel01}.

In this section, we use our mock catalogues to test whether a model in
which AGN  are preferentially  located at the  centres of  dark matter
halos can fit our observational results. As mentioned in section 2, we
have  used  the methodology  introduced  by  \citet{wang06} to  assign
stellar  masses  to the  galaxies  in  the  catalogues. These  authors
adopted  parametrized functions  to relate  galaxy properties  such as
stellar mass to the quantity  $M_{infall}$, defined as the mass of the
halo  at the  epoch where  the galaxy  was last  the  central dominant
object in its  own halo.  It was demonstrated  that these parametrized
relations  were  able  to  provide  an  excellent  fit  to  the  basic
statistical properties of galaxies  in the SDSS, including the stellar
mass function and the shape and amplitude of the two-point correlation
function function  evaluated in different stellar mass  ranges. We now
introduce  a simple  model in  which $p_{AGN}$,  the probability  of a
galaxy to be  an AGN depends only on whether it  is the central galaxy
of its own halo.

In order to create mock AGN and control catalogues that we can compare
directly  with  the  observational   data,  we  follow  the  following
procedure. For  every AGN in our  sample, we select  galaxies from the
mock catalogue that have the  same stellar mass and the same redshift.
We then choose an AGN from  among these galaxies based on whether they
are central  or satellite systems.  The control  galaxies are selected
at random  from the same  set.  The AGN  and control samples  are then
cross-correlated with a  reference sample that is drawn  from the mock
catalogue in exactly  the same way as our  real SDSS reference sample.
The  top  panel  in  Figure  10 shows  how  the  AGN/reference  galaxy
cross-correlation function  changes as a  function of the  fraction of
AGN that are  central galaxies. Note that if  the probability of being
an AGN  is {\em independent} of whether  the galaxy is a  central or a
satellite system, 73\% of the AGN will be central galaxies. The bottom
panel of  Figure 10 shows  how the ratio  between the AGN  and control
galaxy cross-correlation functions varies as this fraction changes.

As  the fraction of  centrally-located AGN  increases, the  ``dip'' on
scales smaller than 1 Mpc  becomes more and more pronounced.  There is
also  a  small  decrease  of the  ratio  $w^{\rm  AGN/ref}(r_p)/w^{\rm
control/ref}(r_p)$ on large scales.  The latter effect arises because,
as more and more AGN are  required to be central galaxies in their own
halos, they also shift into lower mass halos. High mass halos are less
abundant than  low mass  halos and by  definition, each halo  can only
contain one  central galaxy.  As  the central galaxy criterion  on AGN
becomes   more  stringent,  fewer   AGN  will   reside  in   halos  of
$10^{14}-10^{15} M_{\odot}$  and more in halos  with $10^{12} -10^{13}
M_{\odot}$.  The biggest effect, however,  the dip on scales less than
1 Mpc, results from the fact that central galaxies in lower mass halos
have fewer neighbours than non-central galaxies within rich groups.

In Figure  11, we  compare the  results of our  simple model  with the
observational data. In the top  panel, the solid and open circles show
the  cross-correlation functions of  the AGN  and the  control samples
respectively. The solid and dashed  lines show our best-fit models, in
which 84\%  of all AGN  are located at  the centres of their  own dark
matter halos.   In the bottom panel  we compare the  ratio of $w(r_p)$
for the observed  AGN and control galaxies with  that obtained for the
model. The error  bars plotted on the model  curve provide an estimate
of the uncertainty in the result due to {\em cosmic variance effects}.
In order to  estimate these errors, we have  created 20 different mock
catalogues by repositioning the  virtual observer at random within the
simulation   volume.   For   each  mock   catalogue,  we   repeat  our
computations   of  the  AGN   and  control   galaxy  cross-correlation
functions.  The error  bars  are  then calculated  by  looking at  the
variance in the  ratio $w^{\rm AGN/ref}(r_p)/w^{\rm control/ref}(r_p)$
for AGN  and control galaxies selected from  different catalogues. The
errors  estimated  in this  way  are similar  in  size  to the  errors
estimated  by calculating the  variance in  $w^{\rm control/ref}(r_p)$
from  different mock  catalogues; these  errors are  indicated  by the
dashed red lines  in Figure 11.  As can be seen,  the model provides a
good  fit to  the observations  from scales  of $\sim  30$ kpc  out to
scales beyond 10  Mpc.  On scales smaller than 30 kpc,  the AGN show a
small, but significant excess in clustering with respect to the model.
This  is nicely in  line with  the results  presented in  the previous
section.

\section {Summary and Conclusions}

In this paper,  we have analyzed the clustering  of Type 2 narrow-line
AGN  in the  local  Universe using  data  from the  Sloan Digital  Sky
Survey.  The two physical questions we  wish to address are, a) How do
the locations of galaxies  within the large-scale distribution of dark
matter  influence ongoing  accretion onto  their central  black holes?
b)Is  AGN activity triggered by  interactions and  mergers between
galaxies?  To answer these  questions, we analyze the scale-dependence
of  the  AGN/galaxy  cross-correlation  function relative  to  control
samples of non-AGN that are closely matched in stellar mass, redshift,
structural properties,  and mean stellar  age as measured by  the 4000
\AA\  break  strength.   This  close  matching  is  important  because
previous work has established  that the clustering of galaxies depends
strongly  on  properties such  as  luminosity,  stellar mass,  colour,
spectral  type, mean  stellar age,  concentration and  stellar surface
mass density  \citep{norberg02,zehavi02,zehavi05,li06}.  Previous work
has also  established that AGN are  not a random subsample  of the the
underlying  galaxy population.   Rather,  they are  found in  massive,
bulge-dominated galaxies; powerful AGN  tend to occur in galaxies with
smaller black  holes and younger-than-average  stellar populations for
their  mass \citep{kauffmann03,heckman04}.  If  we wish  to understand
whether there is a real  physical connection between the location of a
galaxy  and the  accretion  state of  its  central black  hole, it  is
important that we normalize out these zero'th order trends with galaxy
mass, structure and mean stellar age.

When we  compare the clustering  of AGN relative to  carefully matched
control samples,  and we take the  errors due to  cosmic variance into
account, we obtain the following results:
\begin {enumerate}
\item On scales larger than a few Mpc, the clustering amplitude of AGN
      hosts  does not differ  significantly from  that of  similar but
      inactive galaxies.
\item On  scales between 100  kpc and 1  Mpc, AGN hosts  are clustered
      more  weakly  than  control  samples  of  similar  but  inactive
      galaxies.
\item  On scales  less than  70 kpc,  AGN cluster  more  strongly than
      inactive galaxies, but the effect  is weak. The excess number of
      close companions is only one per hundred AGN.
\end {enumerate}

Our  clustering results  on  large scales  demonstrate  that the  host
galaxies of  AGN are  found in similar  dark matter halos  to inactive
galaxies with  the same structural properties and  stellar masses.  We
have  used  mock catalogues  constructed  from high-resolution  N-body
simulations to show that the AGN anti-bias on scales between 0.1 and 1
Mpc can be explained by  AGN residing preferentially at the centres of
their dark  matter halos.  Our  result on small scales  indicates that
although interactions  may be responsible for  triggering AGN activity
in a  minority of  galaxies, an alternative  mechanism is  required to
explain the nuclear activity in the majority of these systems.

As we have already mentioned, it is easy to understand why dark matter
halo centres  may be preferential  places for ongoing growth  of black
holes. These are  the regions where gas would be  expected to cool and
settle through  radiative processes.  In  addition, dynamical friction
will erode the orbits of  satellite galaxies within a dark matter halo
until they sink to the middle and merge with the central object.  Both
these processes  may bring  fresh gas to  the central galaxy  and fuel
episodes of nuclear activity and  black hole growth.  As we have seen,
however, the evidence for an  excess number of close neighbours around
AGN  is rather  weak,  perhaps  because in  most  cases the  offending
satellite has already been swallowed.  We also note that even the most
powerful AGN  in our  sample are less  luminous than the  quasars with
$M(i) < -24$  for which \citet{serber06} detected an  excess number of
companions on small scales.  What about the evidence for cooling?

Direct observational evidence for  cooling from hot X-ray emitting gas
at   the    centers   of   dark    matter   halos   has    also   been
elusive. \citet{benson00}  used ROSAT PSPC data to  seach for extended
X-ray  emission  from  the  halos  of  three  nearby,  massive  spiral
galaxies.   Their 95  percent  upper limits  on  the bolometric  X-ray
luminosities of the halos show that the present day accretion from any
hot virialized  gas surrounding the galaxies is  very small.  Recently
\citet{pedersen06} detected a gaseous halo aroung the quiescent spiral
NGC 5746 using Chandra observations,  but this remains the only spiral
galaxy with evidence for  ongoing accretion from an extended reservoir
of hot  gas.  In clusters, X-ray  spectroscopy has shown  that most of
the   gas   gas   does   not   manage   to   cool   below   $10^7$   K
\citep[e.g.][]{david01,peterson01}.

Gas accretion in the form of cold HI-emitting clouds is, however, much
less well-constrained.  In  recent work, \citet{kauffmann06} studied a
volume-limited sample of bulge-dominated  galaxies with data both from
the Sloan  Digital Sky Survey  and from the Galaxy  Evolution Explorer
(GALEX)  satellite.   Almost   all  galaxies  with  bluer-than-average
NUV-$r$  colours were  found to  be AGN.   By analyzing  GALEX images,
these authors demonstrated  that the excess UV light  is nearly always
associated  with an extended  disk.  They  then went  on to  study the
relation between the UV-bright outer  disk and the nuclear activity in
these galaxies.  The data indicate  that the presence of the UV-bright
disk  is a  necessary  but  not sufficient  condition  for strong  AGN
activity in a galaxy.  They suggest that the disk provides a reservoir
of fuel for  the black hole. From time to  time, some event transports
gas to the nucleus, thereby triggering the observed AGN activity.

The GALEX results indicate that the extended disks of galaxies play an
important role in the fuelling of AGN. The clustering results from the
SDSS indicate  that AGN are  preferentially located at the  centres of
dark  matter  halos.  In  theoretical  models, rotationally  supported
disks are {\em  expected} to form at the centers  of dark matter halos
\citep{mmw98}.  After the galaxy is accreted by more massive halos and
becomes a satellite system, the disks may lose their gas via processes
such  as   ram-pressure  stripping  \citep[e.g.][]{cayatte94}.   Disks
located  at halo  centres  are  likely likely  to  survive for  longer
periods.  Dynamical perturbations driven  by the dark mattter near the
centres of  the halos may  result in gas  inflows and fuelling  of the
central  black   hole  \citep[see][for  a   recent  discussion]{gw06}.
Further  progress in  understanding the  AGN phenomenon  in  the local
Universe will require detailed  modelling of the observable components
of  galaxies within  evolving dark  matter halos,  as well  as further
investigation  of the  connection between  AGN activity  and phenomena
such as bars, warps, lopsided images, and asymmetric rotation curves.

\section*{Acknowledgments}

We thank the referee for helpful comments.
CL acknowledges the financial  support by the exchange program between
Chinese Academy of Sciences and the Max Planck Society.

The Millennium Run simulation used in this paper was carried out by the 
Virgo Supercomputing Consortium at the Computing Centre of the Max-Planck 
Society in Garching. The semi-analytic galaxy catalogue is publicly 
available at http://www.mpa-garching.mpg.de/galform/agnpaper

Funding for  the SDSS and SDSS-II  has been provided by  the Alfred P.
Sloan Foundation, the Participating Institutions, the National Science
Foundation, the  U.S.  Department of Energy,  the National Aeronautics
and Space Administration, the  Japanese Monbukagakusho, the Max Planck
Society, and  the Higher Education  Funding Council for  England.  The
SDSS Web  Site is  http://www.sdss.org/.  The SDSS  is managed  by the
Astrophysical    Research    Consortium    for    the    Participating
Institutions. The  Participating Institutions are  the American Museum
of  Natural History,  Astrophysical Institute  Potsdam,  University of
Basel,   Cambridge  University,   Case  Western   Reserve  University,
University of Chicago, Drexel  University, Fermilab, the Institute for
Advanced   Study,  the  Japan   Participation  Group,   Johns  Hopkins
University, the  Joint Institute  for Nuclear Astrophysics,  the Kavli
Institute  for   Particle  Astrophysics  and   Cosmology,  the  Korean
Scientist Group, the Chinese  Academy of Sciences (LAMOST), Los Alamos
National  Laboratory, the  Max-Planck-Institute for  Astronomy (MPIA),
the  Max-Planck-Institute  for Astrophysics  (MPA),  New Mexico  State
University,   Ohio  State   University,   University  of   Pittsburgh,
University  of  Portsmouth, Princeton  University,  the United  States
Naval Observatory, and the University of Washington.


\bsp
\label{lastpage}
                                                                                
\end{document}